\documentclass[10pt,oneside,final]{IEEEtran}
\usepackage{amssymb}
\usepackage{amsmath}

\usepackage{graphicx}
\usepackage{color}
\usepackage{multicol}

\usepackage{subfigure}
\usepackage{bm}
\usepackage{latexsym}
\usepackage{stfloats}
\usepackage{float}
\usepackage{exscale}
\usepackage{relsize}
\usepackage{cite}
\usepackage{cases}
\usepackage{algorithm}
\usepackage{algpseudocode}
\usepackage{amsthm}
\usepackage{epsfig}
\usepackage{epstopdf}
\usepackage{breqn}
\usepackage{enumerate}
\usepackage{multirow}
\usepackage[utf8]{inputenc}
\usepackage{algorithmicx}
\usepackage{amssymb}

\begin{document}
	\title{  \vspace{-2.4ex} Deep Learning-Based CSI Feedback for XL-MIMO Systems in the Near-Field Domain}
	\vspace{-5ex}
	\author{
		Zhangjie Peng,  Ruijing Liu, Zhaotian Li,
		Cunhua Pan,~\IEEEmembership{Senior Member,~IEEE},
		and Jiangzhou Wang,~\IEEEmembership{Fellow,~IEEE}\vspace{-6ex}

		\thanks{Z. Peng, R. Liu and Z. Li are with the College of Information, Mechanical and Electrical Engineering, Shanghai Normal University, Shanghai 200234, China. (e-mail: pengzhangjie@shnu.edu.cn; 1000511821@smail.shnu.edu.cn; 1000511820@smail.shnu.edu.cn;).}
		\thanks{C. Pan is with the National Mobile Communications Research Laboratory, Southeast University, Nanjing 210096, China. (e-mail: cpan@seu.edu.cn).}
		\thanks{J. Wang is with the School of Engineering, University of Kent, CT2 7NT Canterbury, U.K. (e-mail: j.z.wang@kent.ac.uk).}

	}
	\maketitle
	\newtheorem{lemma}{Lemma}
	\newtheorem{theorem}{Theorem}
	\newtheorem{remark}{Remark}
	\newtheorem{corollary}{Corollary}
	\newtheorem{proposition}{Proposition}

	\begin{abstract} 
		 
		 In this paper, we consider an extremely large-scale massive multiple-input-multiple-output (XL-MIMO) system. As the scale of antenna arrays increases, the range of near-field communications also expands. In this case, the signals no longer exhibit planar wave characteristics but spherical wave characteristics in the near-field channel, which makes the channel state information (CSI)  highly complex. Additionally, the increase of the antenna arrays scale also makes the size of the CSI matrix significantly increase. Therefore, CSI feedback in the near-field channel becomes highly challenging. To solve this issue, we propose a deep-learning (DL)-based ExtendNLNet that can compress the CSI, and further reduce the overhead of CSI feedback. In addition, we have introduced the Non-Local block to obtain a larger area of CSI features. Simulation results show that the proposed ExtendNLNet can significantly improve the CSI recovery quality compared to other DL-based methods.
	\end{abstract}

	\begin{IEEEkeywords}
		Deep Learning, XL-MIMO, near-field domain, CSI feedback.
	\end{IEEEkeywords}

	\section{Introduction}

As one of the key technologies for 6G, extremely large-scale massive multiple-input-multiple-output (XL-MIMO) systems  have a larger number of antennas than massive MIMO systems and can achieve high spectral efficiency and high energy efficiency \cite{1}. However, the Rayleigh distance  increases as the number of antennas increases, leading to an expansion of the range of near-field communications \cite{2,19,20}. Therefore, the research on near-field communications will become essential in studying XL-MIMO systems.

For near-field communications, the authors of \cite{3} considered the performance analysis of XL-MIMO systems in the near-field domain, including the signal-to-noise ratio (SNR) scaling laws and achievable data rate. In \cite{4}, the authors proposed an algorithm to achieve low-complexity near-field channel estimation for XL-MIMO systems in the near-field domain. 
In \cite{12}, the authors used beamspace modulation to improve the capacity by exploiting the increased degrees of freedom in the near-field domain of XL-MIMO systems.
In \cite{13}, the authors proposed an efficient beam alignment (BA) algorithm for XL-MIMO systems in the near-field domain, which can achieve low BA error rate and overhead.

As a new direction of machine learning, deep learning (DL) has been widely applied in the communication field to improve system performance \cite{21,22}.
The authors of \cite{5} proposed a DL-based network to estimate the multi-user high-dimensional channel in XL-MIMO systems for both near-field and far-field users. The authors of \cite{6} proposed a DL-based beamforming training scheme for XL-MIMO systems, and used supplementary codewords to enhance the beamforming training. Similarly, the authors of \cite{7} also proposed a DL based beamforming training scheme for XL-MIMO systems in the near-field domain, considered the near-field codebook, and improved the performance of beam training.

Different with time division duplex (TDD) systems, the uplink and downlink channels in frequency division duplexing (FDD) systems lack channel reciprocity. Therefore, channel estimation must be completed at the user, and the user needs to feedback channel state information (CSI) to the base station (BS) in FDD systems \cite{8}. Since the increase of the antenna array scale also makes the size of the CSI matrix significantly increase, feedback the CSI matrix directly would occupy a significant portion of channel bandwidth resources. Therefore, CSI needs to be compressed and restored during the feedback process. The authors of \cite{9} proposed a codebook-based method to compress CSI in massive MIMO systems. In \cite{10}, a method based on compressive sensing was proposed to compress CSI in massive MIMO systems. 
The authors of \cite{14}  proposed a multiple-rate compressive sensing neural network framework to compress and quantize the CSI.
However, these methods are not satisfactory in term of the accuracy of the CSI decompression, and the research on CSI feedback for XL-MIMO systems in the near-field domain is still not studied.

In this paper, we consider the CSI feedback for XL-MIMO systems in the near-field domain, and the contributions are summarized as follows: 1) A network called ExtendNLNet is proposed for XL-MIMO systems to compress and decompress the CSI of the near-field channel; 2) We introduce the Non-Local block to obtain a larger scale of CSI features; 3) Simulation results demonstrate that the performance of the proposed ExtendNLNet improves significantly as the compression ratio increases.

\section{System Model}

\begin{figure}[t]
	\centering
	\includegraphics[width=0.65\linewidth]{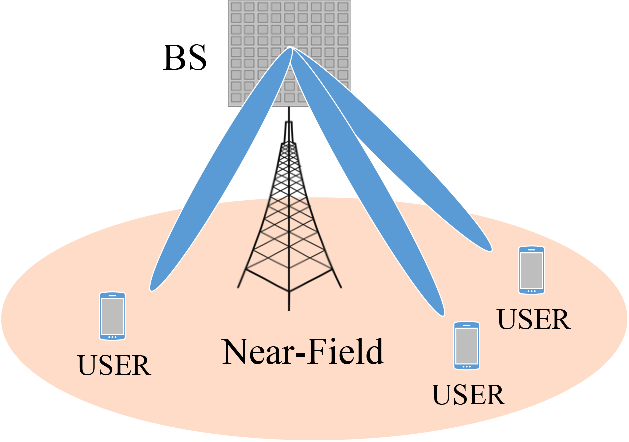}\vspace{-0.3cm}
	\caption{{Near-field communications of XL-MIMO system.} } 
	\label{fig1}\vspace{-3.5ex}
\end{figure}

\subsection{Signal Model}

As shown in Fig. \ref{fig1}, we consider an XL-MIMO system in the near-field domain. The BS is equipped with $N_1$ antennas and the user in the near-field domain is equipped with $N_2$ antennas.  The channel from the BS to the user is denoted by $\textbf{H}\in \mathbb{C}^{N_2\times N_1}$. Therefore, the signal received by the user can be expressed as
\begin{align}\label{signal}
	\textbf{y}=\textbf{H}\textbf{v}x+\textbf{n},
\end{align}
where  $\textbf{v}\in \mathbb{C}^{N_1\times 1}$ denotes the precoding vector, $x$ is the  transmitted signal, and $\textbf{n}\sim \mathcal{CN} (0,\delta^{2} \mathbf{I}_{N_2})$ denotes the additive white Gaussian noise received by the user. 
\vspace{-1ex}
\subsection{Channel Model}

Consider that each BS-user antenna pair experiences different transmission paths in the near-field domain of XL-MIMO systems, we consider the path component for each BS-user antenna pair separately under the assumption of geometric free-space propagation \cite{15}. By defining $\mathbf{H}(n_{2},n_{1})$ as the channel between the $n_{1}$-th antenna at the BS and the $n_{2}$-th antenna at the user, it can be expressed as
\begin{align}\label{JCF}
	\mathbf{H}(n_{2},n_{1})=\frac{1}{r_{n_{2},n_{1}} } e^{-j\frac{2\pi }{\lambda } r_{n_{2},n_{1}}},
\end{align}
where $n_{1}\in  \{1,2,...,N_{1}\}$ and $n_{2}\in \{1,2,...,N_{2}\}$. $r_{n_2,n_1}$ represents the transmission distance from the $n_1$-th antenna at the BS to the $n_2$-th antenna at the user. Additionally,  $\frac{1}{r_{n_{2},n_{1}} }$ represents the normalized free-space path loss of each BS-user antenna pair, and $r_{n_{2},n_{1}}$ can be expressed as
\begin{align}\label{r}
		&r_{n_{2},n_{1}}=\!\sqrt{(r\cos \theta -d_{2} \sin \phi )^{2}+(r\sin \theta+d_{2} \cos \phi-d_{1} )^{2}  } \notag\\
		&\!=\!\sqrt{\! r^{2} \!\!+\! d_{1}^{2}\!+\! d_{2}^{2}\!+\! 2(rd_{2}\!\sin(\theta \!\!+\!\!\phi )\!-\! rd_{1}\!\sin \theta \!-\! d_{1}d_{2}\!\cos \phi )},
\end{align}
where $r$ is the distance between the first antenna of the user and the first antenna of the BS, $\phi$ represents the relative angle between the BS and the user, and $\theta$ represents the transmission angle of the signal. Additionally, $d_{1}$ and $d_{2}$ are respectively expressed as
\begin{align}\label{bg}
	&d_{1}=n_{1}d,\\
	&d_{2}=n_{2}d,
\end{align}
where $d$ is the antenna spacing. Therefore, we can obtain the channel model as
\begin{align}\label{JCF2}
	\mathbf{H}(r,\theta, \phi)=\left[ \frac{1}{r_{n_{2},n_{1}} } e^{-j2\pi\frac{r_{n_{2},n_{1}} }{\lambda }} \right]_{N_{2}\times N_{1}}.
\end{align}

\section{CSI Feedback Process and ExtendNLNet}
\subsection{CSI feedback process}

After the user estimates the complete CSI matrix $\mathbf{H}$, it needs to be split into real and imaginary parts because neural networks are not suitable for processing complex numbers. This split operation is performed before feeding the matrix into the neural network and the resulting split matrix is denoted as $\mathbf{H}_{\textrm{in}}$. As the original CSI matrix $\mathbf{H}$ is of size $N_2 \times N_1$, the split matrix $\mathbf{H}_{\textrm{in}}$ will then have a size of $L = 2N_2 N_1$. After splitting, each element in the matrix $\mathbf{H}_{\textrm{in}}$ is normalized within $[0,1]$. Then, the encoder part at the user will compress the $L$-sized matrix $\mathbf{H}_{\textrm{in}}$ into a $K$-dimensional feature vector ${\tilde{\textbf s}}$ based on the given compression ratio ($C \! R$). The $C \! R$ is expressed as
\begin{align}\label{CR}
	C\!R={L}/{K}.
\end{align}

The compression process can be expressed as
\begin{align}\label{com}
	&\tilde{\textbf s}=f_{\textrm{en}}\left ( \textbf{H}_{\textrm{in}},\theta _{\textrm{en}} \right ),
\end{align}
where $f_{\textrm{en}}\left(\cdot\right)$ is the function used for the compression process, and $\theta_{\textrm{en}}$ represents the parameters of the encoder.

Once the BS receives the feature vector $\tilde{\textbf s}$, the decoder at the BS decompresses the $K$-dimensional feature vector into an $L$-sized CSI matrix $\textbf{H}_{\textrm{out}}$, and the decompression process can be expressed as
\begin{align}\label{decom}
	&\textbf{H}_{\textrm{out}}=f_{\textrm{de}}\left ( \tilde{\textbf s},\theta _{\textrm{de}} \right ),
\end{align}
where $f_{\textrm{de}}\left ( \cdot  \right )$ is the decompression function, and $\theta _{\textrm{de}}$ represents the parameters of the decoder.
In addition, we use the Mean Square Error (MSE) as the loss function. By substituting \eqref{com} and \eqref{decom} into the loss function, the optimized expression of the neural network can be obtained as
\begin{align}\label{12}
	&\left ( {\hat{\theta}}_{\textrm{en}},{\hat{\theta}}_{\textrm{de}} \right )=\mathop{\arg\min\limits_{\theta _{\textrm{en}},\theta _{\textrm{de}}}}\left\| \textbf{H}_{\textrm{out}}-\textbf{H}_{\textrm{in}}\right\|_{2}^{2},
\end{align}
where ${\hat{\theta}}_{\textrm{en}}$ and ${\hat{\theta}}_{\textrm{de}}$ represent the  optimal parameters of the $f_{\textrm{en}}\left(\cdot\right)$ and $f_{\textrm{de}}\left(\cdot\right)$, respectively.

\subsection{Architecture of ExtendNLNet}
\begin{figure*}[ht]	
	\centering	\vspace{-4ex}
	\includegraphics[height=\linewidth,angle=90]{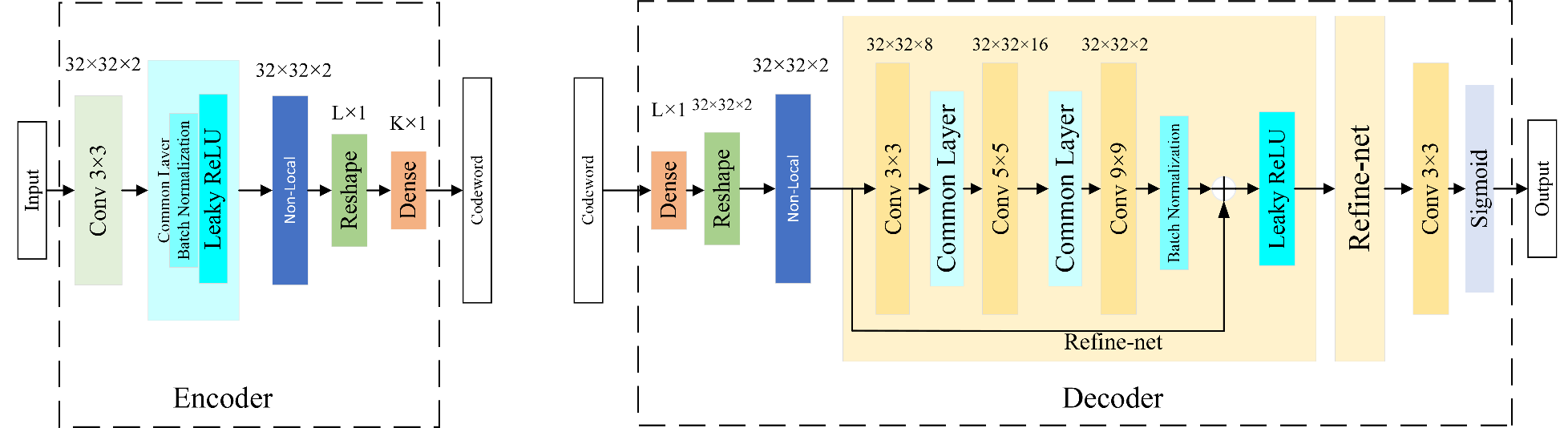}	
	\vspace{-3ex}
	\caption{The architecture of ExtendNLNet.}	
	\label{fig2}
	\vspace{-3ex}
\end{figure*}
As shown in Fig. \ref{fig2}, we show the architecture of our proposed ExtendNLNet. Specifically, after the CSI matrix is input into the encoder, it first undergoes convolution process with a $3\times 3$ convolution layer. Then, the convolution result is batch normalized, and we use a LeakyReLU   \cite{17} layer to perform the non-linear activation process. After the activation process, the data is input into the Non-Local block for further feature learning and information extraction. The specific structure and design principles of the Non-Local block will be discussed in details. For the output from the Non-Local block, the CSI matrix is reshaped into an $L\times 1$ vector and then compressed. 

After the codeword obtained by the decoder at the BS, it first executes the reverse process of the encoder. Specifically, the codeword is remapped into an $L \times 1$ vector by a fully connected layer, and then reshaped into two $32 \times 32$ matrices. These two matrices serve as the initial results for the real part and imaginary part of the CSI matrix. Next, the initial results are input into the Non-Local block for feature learning, and the CSI matrix is preliminarily recovered. Then, the results from the Non-Local block are input into the Refine-net  block \cite{16}, which continuously refine the reconstruction to make the initial results more consistent with the input matrix .

In ExtendNLNet, the three convolutional layers of the Refine-net block do not all use  $3 \times 3$ sized convolutional layer. 
Considering a single sized convolutional layer may not adapt the changes in the sparsity of the CSI matrix well, which may lead to a decrease in the performance of the neural network. Meanwhile, different sized convolutional layer have different receptive fields, which can adapt to different sparsity levels. Specifically, convolutional layers with smaller size can better extract finer features, suitable for information-dense areas, and larger size is better for relatively sparse parts.
Therefore, the first convolutional layer is of size $3 \times 3$, and the second convolutional layer is of size $5 \times 5$. They respectively expand the CSI matrix to 8 and 16 feature maps. The size of the third convolutional layer is $9 \times 9$, which restores the number of feature maps to 2. 
Moreover, batch normalization is applied after each convolution layer, and using the LeakyReLU for activation at last. In addition, the data from the initial input of the Refine-net block is added back to avoid the problem of gradient vanishing.

After passing through two Refine-net blocks, the data undergoes another $3 \times 3$ sized convolutional layer, activated by a Sigmoid layer, and output in the end. During the training of the neural network, the network computes the input and output results into the loss function, aiming to iteratively update its parameters to achieve lower values of the loss function.

\subsection{Non-Local Block}
Existing researches of DL-based CSI feedback architectures typically use convolutional or recurrent operations to extract features, which can only handle a small area of the CSI matrix at a time. These operations need to be repeated to obtain a larger area of the CSI matrix, which leads to algorithm optimization difficulties and computational inefficiency. Therefore, we use the Non-Local \cite{18} block to extract CSI matrix features over a large range. The Non-Local block is designed specifically for sequential data and can directly capture the feature relationships between any two positions in the CSI image. The calculation is formulated as
\begin{align}\label{nl}
	u( \mathbf{a}_{i} )=\sum_{\mathbf{\tilde{b}}\in\Omega}^{}w( \mathbf{a}_{i},\mathbf{\tilde{b}} )v( \mathbf{\tilde{b}} ),
\end{align}
where $w(\mathbf{a}_{i}, \mathbf{\tilde{b}})$ is the normalized weight and $i$ represents the index of the position on the feature map. The $\mathbf{\tilde{b}}$ that is more similar to the output result $u(\mathbf{a}_{i})$ will have larger weight in the calculation. Similar to \eqref{nl}, the general Non-Local process can be expressed as
\begin{align}\label{nl2}
	\mathbf{y}_{i}=\frac{1}{C(\mathbf{x})}\sum_{\forall j}^{}f( \mathbf{x}_{i},\mathbf{x}_{j} )g(\mathbf{x}_{j} ),
\end{align}
where $C(\mathbf{x})$ is the normalization factor. $\mathbf{x}$  and $\mathbf{y}$ represent  the input and output feature maps, respectively.  $j$ represents the index of all possible positions on $\mathbf{x}$. The function $g$ is used to compute the embedded feature representation of the input feature map at position $j$. The function $f$ is used to compute the correlation between indices $i$ and $j$.  In this paper, we use the embedded Gaussian as function $f$, and can be expressed as
\begin{align}\label{nl3}
	f( \mathbf{x}_{i},\mathbf{x}_{j} )=e^{\tilde{\theta}(\mathbf{x}_{i})^{T}\tilde{\phi}(\mathbf{x}_{j})},
\end{align}
where $\tilde{\theta}$ and $\tilde{\phi}$ represent embedding spaces.

\begin{figure}[h]
	\vspace{-3ex}
	\centering
	\includegraphics[scale=0.63]{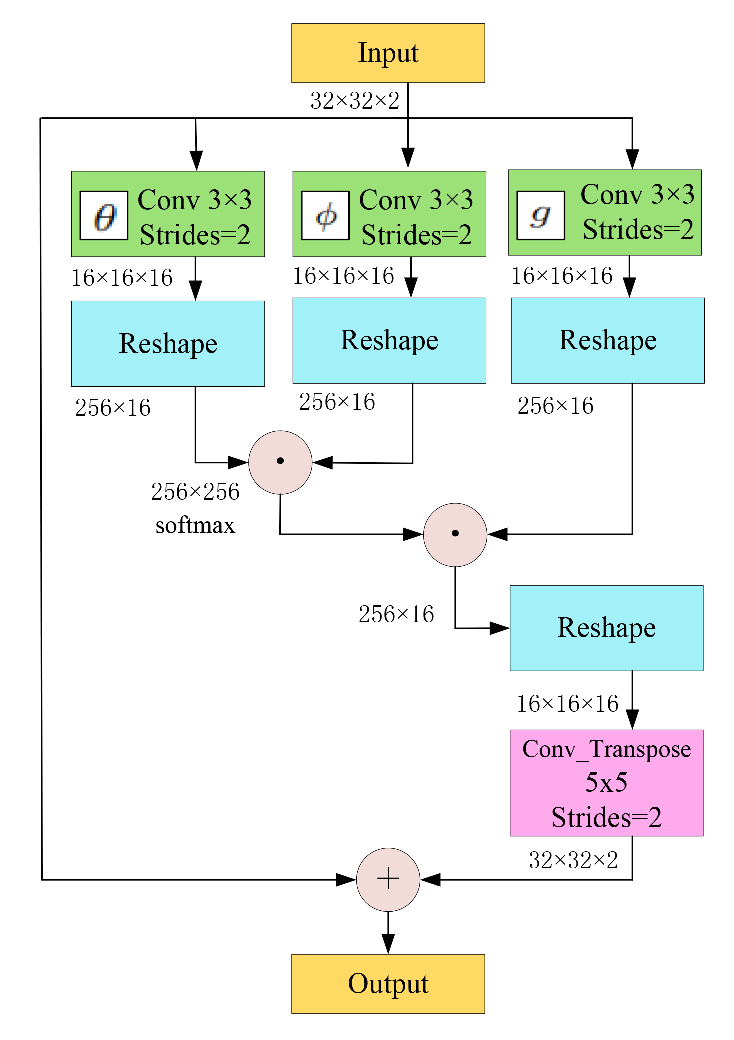}\vspace{-2ex}
	\caption{The architecture of Non-Local block.} \label{fig3}\vspace{-2ex}
\end{figure}

The architecture of the Non-Local block is illustrated in Fig. \ref{fig3}. The overall operation of the entire block can be expressed as
\begin{align}\label{nl4}
	\mathbf{z}_{i}=NL(\mathbf{y}_{i})+\mathbf{x}_{i},
\end{align}
where $NL(\cdot)$ represents the Non-Local block operation, $\mathbf{y}_{i}$ is given in \eqref{nl2} and $+ \mathbf{x}_i$ represents the residual connection. To reduce the size of the feature maps in the Non-Local block, convolution operations with a stride of 2 and a kernel size of $3 \times 3$ are added as downsampling operations. In addition, this reduces the size of the multiplied matrices from the original $1024 \times 1024$ to $256 \times 256$. Simultaneously, the number of channels is increased to 16 to compensate for the loss in neural network performance caused by downsampling. Before adding with the residual connection, a transposed  $3 \times 3$ sized convolution and a stride of 2 is used as an upsampling operation to restore the shape of the feature maps. 

As shown in \eqref{nl2}, the Non-Local operation calculates the correlation $f(\mathbf{x}_i, \mathbf{x}_j)$ at all positions, enabling it to directly extract and propagate correlated features between position $i$ and position $j$ in one operation, which can be regarded as an operation that  provide a global view for feature extraction. Specifically, the global convolution covering the entire feature map with its own self-correlation matrix as the kernel. It effectively captures the correlation between two distant positions, making the Non-Local block more suitable for holistic feature extraction. Consequently, optimal results can be achieved without introducing too many parameters.

\section{Simulation Results}
To provide numerical results, the network complexity analysis is shown in Table \ref{tab4:fz}, and we use the number of parameters to express the complexity of the networks. As shown in Table \ref{tab4:fz}, by comparing the number of parameters between CsiNet, Attention-CSiNet and ExtendNLNet under the same ${C \! R}$, the complexity of the ExtendNLNet has not increased much compared with the other methods.

\begin{table}[t]
	\vspace{-6ex}
	\begin{center}
		\caption{The number of parameters versus $C\!R$}
		\label{tab4:fz}
		\renewcommand\arraystretch{1.5}
		\setlength{\tabcolsep}{1.5mm}{
			\begin{tabular}{|c|c|c|c|c|} 
				\hline
				$C\!R$ & 16 & 32 & 64\\
				\hline
				CsiNet & 530,656 & 268,448 & 137,344\\
				\hline
				Attention-CsiNet & 530,680 & 268,472 & 137,368\\
				\hline
				\textbf{ExtendNLNet} & \textbf{543,456} & \textbf{281,248} & \textbf{150,144}\\
				\hline
			\end{tabular}	}
	\end{center}
\end{table}

\begin{table}[t]
	\vspace{-3ex}
	\begin{center}
		\caption{The performance of different net versus $C\!R$}
		\label{tab:dfCR}
		\renewcommand\arraystretch{1.5}
		\begin{tabular}{|c|c|c|c|c|c|} 
			\hline
			& $C\!R$ & 16 & 32 & 64\\
			\hline
			\multirow{3.5}{0.1cm}{\rotatebox{90}{NMSE}} & CsiNet  & -22.11 & -17.14 & -11.82\\
			\cline{2-5}
			& Attention-CsiNet  & -20.65 & -17.23 & -12.21\\
			\cline{2-5}
			& \textbf{ExtendNLNet}  & \textbf{-22.94} & \textbf{-21.18} & \textbf{-16.77}\\
			\hline
			\multirow{3}{0.1cm}{$\rho$} & CsiNet  & 96.80$\%$ & 93.05$\%$ & 84.92$\%$\\
			\cline{2-5}
			& Attention-CsiNet  & 96.08$\%$ & 93.85$\%$ & 86.18$\%$\\
			\cline{2-5}
			& \textbf{ExtendNLNet}  & \textbf{97.33}$\%$ & \textbf{96.89}$\%$ & \textbf{94.58}$\%$\\
			\hline
		\end{tabular}
		\vspace{-5.5ex}
	\end{center}
\end{table}

To compare the performance of different networks, we introduce cosine similarity $\rho$ and normalized mean square error (${\rm NMSE}$) as the performance indicators\cite{11}. $\rho$ can be expressed as
\begin{align}\label{rho}
	&\rho =\mathbb{E}\left\{ \frac{1}{N_{c}}\sum_{n=1}^{N_{c}}\frac{\left|\textbf{h}_{\textrm{out}\left ( n \right )}^{H}\cdot \textbf{h}_{\textrm{in}\left ( n \right )} \right|}{\left\| \textbf{h}_{\textrm{out}\left ( n \right )}\right\|_{2}\left\| \textbf{h}_{\textrm{in}\left ( n \right )}\right\|_{2}}\right\},
\end{align}
where $N_{c}$ denotes the number of columns of the matrix, $\textbf{h}_{\textrm{out}\left ( n \right )}$  denotes the column vector of the output  matrix, and $\textbf{h}_{\textrm{in}\left ( n \right )}$ denotes the column vector of the  input matrix. The NMSE can be expressed as
\begin{align}\label{NMSE}
	&\textrm{NMSE}=\textrm{log}\left (\mathbb{E}\left\{ \frac{\left\| \textbf{H}_{\textrm{in}}-\textbf{H}_{\textrm{out}} \right\|_{F}^{2}}{\left\| \textbf{H}_{\textrm{in}}\right\|_{F}^{2}}\right\}\right ).
\end{align}

The number of the antennas equipped by the BS and the user are set as $N_1=1024$ and $N_2=1$, respectively. In the training process of the neural networks, the epoch is set to 100, the batch size is set to 250, and the learning rate is set to 0.001. The number of samples in the training set, test set, and validation set are 25,000, 5,000, and 5,000, respectively. To facilitate comparison of network performance, we have compared the performance of CsiNet, Attention-CsiNet, and ExtendNLNet. The optimal performance of each network is summarized and recorded in Table \ref{tab:dfCR}.

\begin{figure}[t]
	\vspace{-6ex}
	\begin{minipage}[t]{1\linewidth}
		\centering
		\includegraphics[scale=0.42]{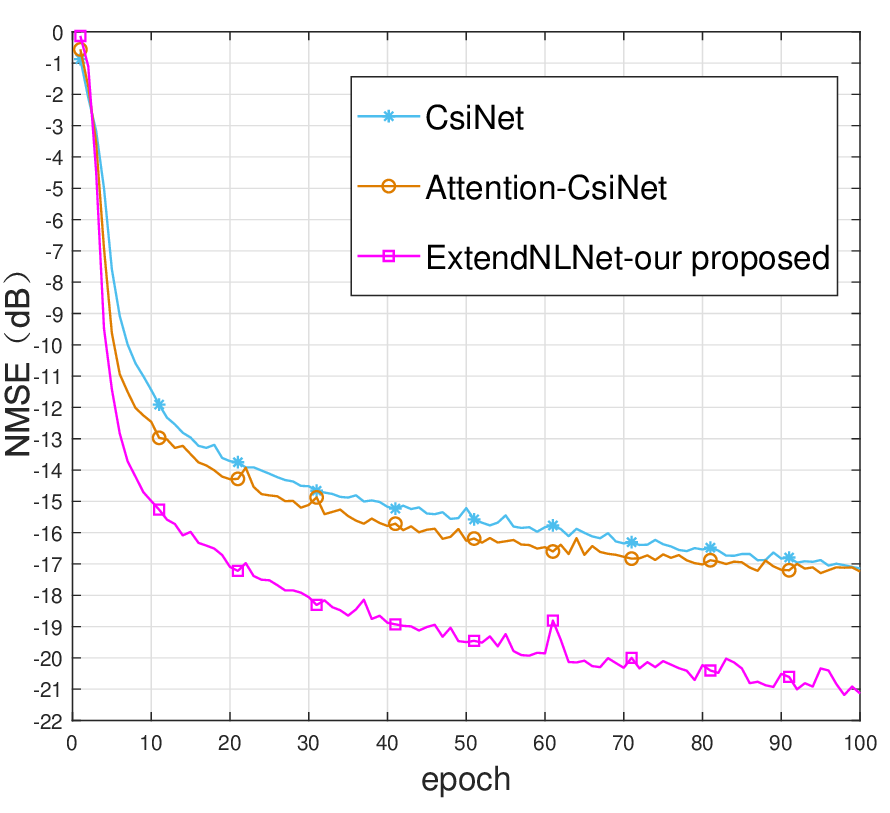}\vspace{-2ex}
		\caption{The ${\rm NMSE}$ when $C\!R=32$. }
		\label{fig4}
	\end{minipage}
	
	\begin{minipage}[t]{1\linewidth}
		\centering
		\includegraphics[scale=0.42]{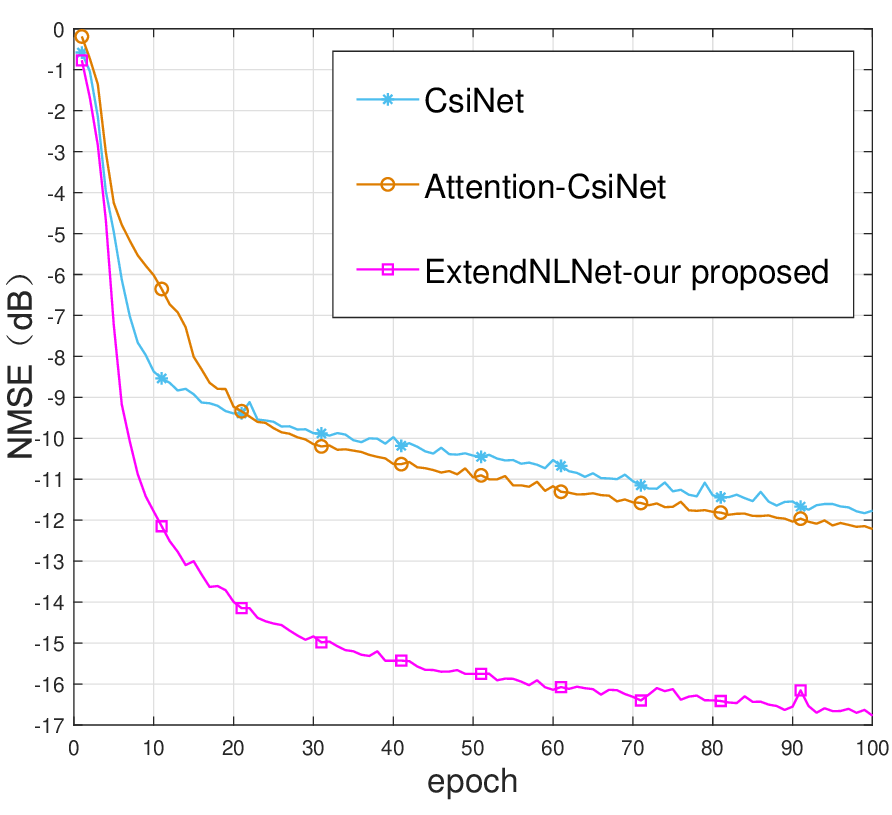}\vspace{-2ex}
		\caption{The ${\rm NMSE}$ when $C\!R=64$.}
		\label{fig5}	\vspace{-3ex}
	\end{minipage}
	\vspace{-4ex}
\end{figure}

As shown in the Table \ref{tab:dfCR}, as the $C \! R$ increases, the performance advantage of ExtendNLNet becomes more prominent. This trend is consistent with the increase of parameters shown in Table \ref{tab4:fz}. Compared to CsiNet, the parameter increment of ExtendNLNet remains relatively consistent across different $C \! R$, while the parameters of CsiNet decrease significantly as the $C \! R$ increases. Meanwhile, ExtendNLNet experiences a larger increase in parameters when the $C \! R$ is high.

Specifically, when the $C \! R$ is 16, the performance improvement of ExtendNLNet compared to CsiNet and Attention-CsiNet is $3.75 \% $ and $11.09\%$, respectively. When the $C \! R$ is 32, the performance improvement of ExtendNLNet compared to CsiNet and Attention-CsiNet is $23.57\%$ and $22.93\%$, respectively. When the $C \! R$ reaches 64, the performance improvement of ExtendNLNet compared to CsiNet and Attention-CsiNet is $41.88\%$ and $37.34\%$, respectively. In contrast, the performance improvement of Attention-CsiNet compared to CsiNet is minimal, and even shows a slight decline when the $C \! R$ is 16. This indicates that for the near-field channels of XL-MIMO communication systems, achieving good CSI matrix recovery requires paying more attention to the global structural characteristics of the CSI matrix, rather than being limited to local features in small blocks. Overemphasizing local features may lead to performance degradation because of being trapped in local optima. The introduction of larger convolutional layers in the Non-Local block and Refine-net block aims to capture large-scale  features. Meanwhile, the small convolutions with kernel sizes of $5\times 5$ and $3\times 3$ in the neural network ensure that the performance of the neural network remains at a high level even at low $C \! R$, and enhancing the robustness and versatility of the neural network.

\begin{figure}[t]

	\begin{minipage}[t]{1\linewidth}
		\centering
		\includegraphics[scale=0.42]{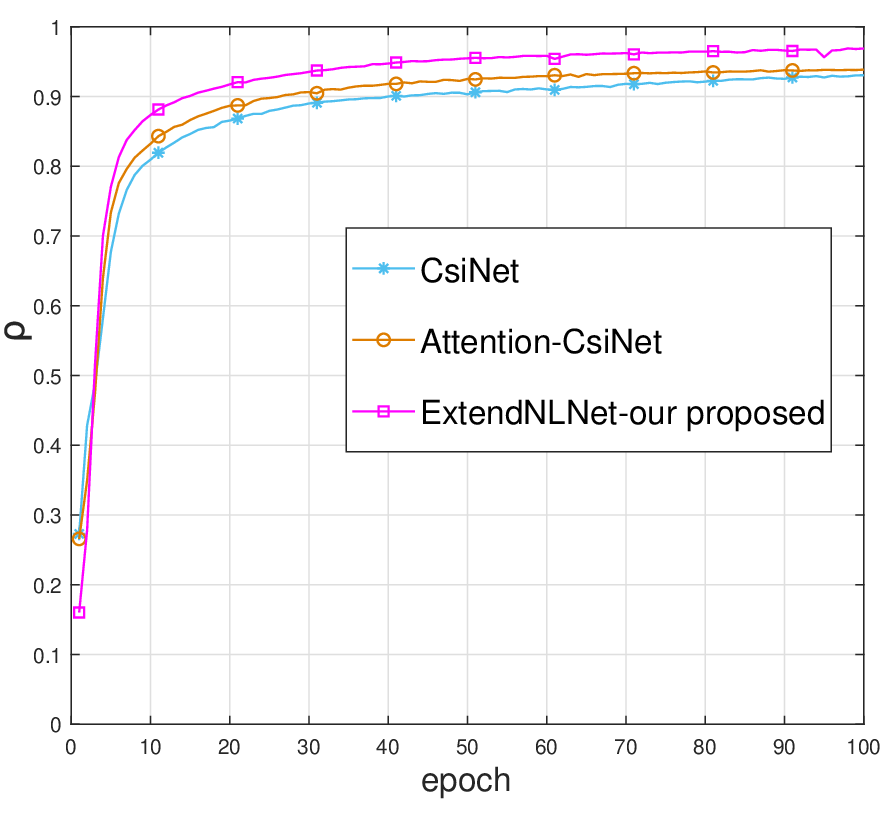}\vspace{-2ex}
		\caption{The $\rho$ when $C\!R=32$.}
		\label{fig6}
	\end{minipage}

	\begin{minipage}[t]{1\linewidth}
		\centering
		\includegraphics[scale=0.42]{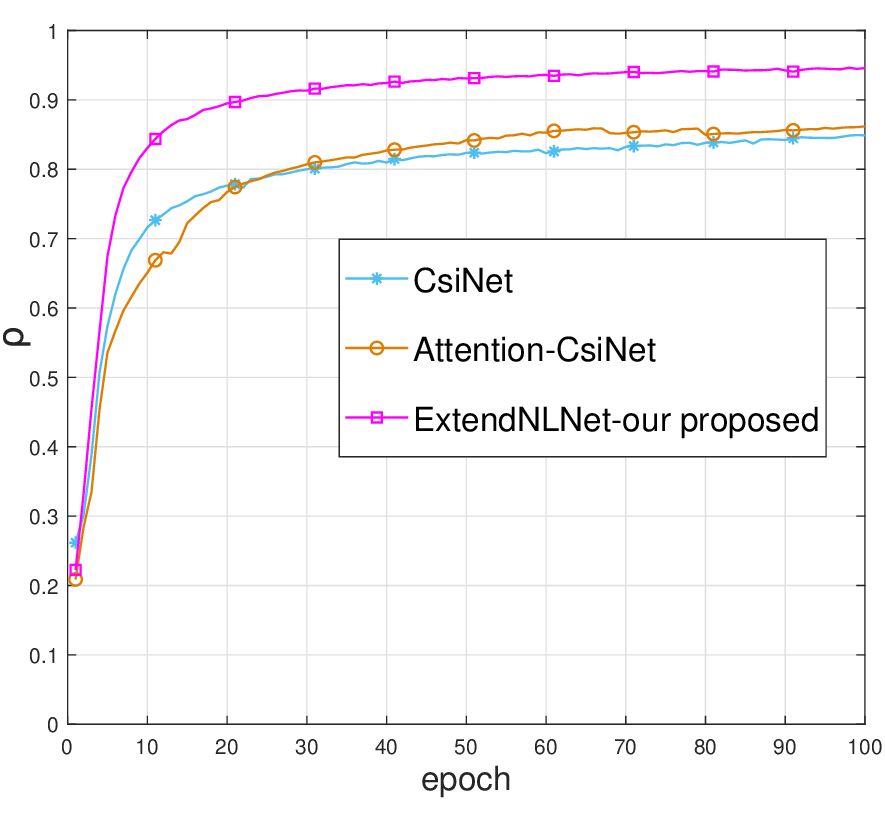}\vspace{-2ex}
		\caption{The $\rho$ when $C\!R=64$.}
		\label{fig7}\vspace{-3ex}
	\end{minipage}
	\vspace{-4ex}
\end{figure}

From Fig. \ref{fig4}, Fig. \ref{fig5}, Fig. \ref{fig6} and Fig. \ref{fig7}, we provide statistics on how the network performance of different neural networks changes during the training process under different $C \! R$, which shows the performance of the neural network improves significantly as the compression ratio increases.

\section{Conclusions}

In this paper, we proposed a DL-based ExtendNLNet for XL-MIMO systems in the near-field domain, which can compress the CSI and further reduce the overhead of CSI feedback. In addition, we introduced the Non-Local block and convolutions with larger kernels to expand the feature extraction field of the proposed ExtendNLNet, which can obtain a larger area of CSI features. Meanwhile, some convolution layers with smaller kernels are retained to maintain  capability of the proposed ExtendNLNet for local feature extraction. Simulation results demonstrated that the performance of the proposed ExtendNLNet improves significantly as the compression ratio increases.

	\bibliographystyle{IEEEtran}
	\bibliography{IEEEabrv,Refer}
	
\end{document}